\documentclass[twocolumn,aps,showpacs,pra,superscriptaddress]{revtex4}
\makeatletter

\newcommand{\Rmnum}[1]{\expandafter\@slowromancap\romannumeral #1@}
\makeatother
\usepackage{amsmath}
\usepackage{graphicx}
\usepackage{subfigure}
\usepackage{color}
\usepackage{amsfonts}
\usepackage{tikz}
\usepackage{esint}
\usepackage{mathrsfs}

\begin{document}
\title{Generalized effective-potential Landau theory for the two-dimensional extended Bose-Hubbard model}

\author{Zhi Lin}
\email{zhilin13@fudan.edu.cn}
\affiliation{School of Physics and Materials Science, Anhui University, Hefei 230601, China}
\affiliation{Department of Physics and Center of Theoretical and Computational Physics, The University of Hong Kong, Pokfulam Road, Hong Kong, China}
\affiliation{Shenzhen Institute of Research and Innovation, The University of Hong Kong, Shenzhen 518063, China}
\affiliation{State Key Laboratory of Surface Physics and Department of Physics, Fudan University, Shanghai 200433, P. R. China}
\author{Ming Yang}
\email{mingyang@ahu.edu.cn}
\affiliation{School of Physics and Materials Science, Anhui University, Hefei 230601, China}

\begin{abstract}
We analytically study the quantum phase diagrams of ultracold dipolar Bose gases in an optical square lattice at zero temperature by using the generalized effective-potential Landau theory (GEPLT). For a weak nearest-neighbor repulsion, our analytical results are better than the third-order strong-coupling expansion theory calculation [M. Iskin et al., \textcolor[rgb]{0.00,0.00,1.00}{ Phys. Rev. A \textbf{79}, 053634 (2009)}].  In contrast to a previous quantum Monte Carlo (QMC) simulation [T. Ohgoe et al., \textcolor[rgb]{0.00,0.00,1.00}{Phys. Rev. B \textbf{86}, 054520 (2012)}], we analytically calculate phase transition boundaries up to the third-order hopping, which are in excellent agreement with QMC  simulations for second-order phase transition.
\end{abstract}
\maketitle
\section{Introduction}
Since realizing the Bose-Hubbard model \cite{BHM1,BHM2} and observing a quantum phase transition from a  Mott insulator (MI) to a superfluid (SF) in optical lattices \cite{Greiner}, ultracold Bose atom systems in optical lattices have been one of hot topics \cite{Lewenstein,RMP}. Various intriguing physics have been realized in such systems, including the Hofstadter model \cite{hofstadter,Measuring-chern-number}, the chiral edge states \cite{chiral-edge-states}, and two-dimensional spin-orbit-coupled systems with topological bands\cite{2D-SOC} etc. These experiments were performed with alkali-metal atoms, where two-body interactions of these atoms are isotropic short-range \textit{s}-wave interactions \cite{RMP} and the effective strength can be controlled by adjusting the intensity of the optical lattice potential or by Feshbach resonances \cite{FR1,FR2,FR3}.

Recently, more and more interests  have shifted from simple systems (single-component alkali-metal atom systems) to complex systems, including multi-component gases \cite{altman1,soltan-panahi1}, frustration phenomenon \cite{eckardt-2010,pielawa,ye-2012},  superlattice structures \cite{piel1,sebby-strabley1,folling1,cheinet1,jo1}, higher bands physics \cite{higher_bands1,higher_bands2}, long-range interactions \cite{lahaye,lauer1,trefzger,naini,schauss1} etc. There are plenty of novel phases \cite{Lewenstein1} in such complex systems, so the corresponding phase diagrams of these systems are abundant and various.

The irreducibly complex systems induced by long-range interactions can be described by the
extended Bose-Hubbard model \cite{trefzger}. It has been studied by numerical methods, involving Gutzwiller ansatz \cite{Gutzwiller,Iskin,Gutzwiller2}, density matrix renormalization group \cite{DMRG} and quantum Monte Carlo (QMC) simulations \cite{QMC1,QMC2}, and analytical methods, including mean-field theory \cite{SC}, random-phase approximation \cite{RPA}, strong-coupling expansion method \cite{SC}, and the generalized Green's function method  \cite{GGF1}. The phase boundaries obtained by random-phase approximation or generalized Green's function method (the lowest order) coincide with the mean field results. Meanwhile, the locations of phase boundaries obtained by strong-coupling perturbation theory always overestimate QMC results. Moreover, in the case of  $ZV\gtrsim 0.7U$ ($Z$ is the coordination, $V$ is nearest-neighbor repulsions number, and $U$ denotes the on-site repulsion), it is necessary to include at least fourth-order correction for using strong-coupling perturbation theory to obtain the phase boundary between the charge-density-wave (CDW) phase and supersolid (SS) phase \cite{SC}, but the analytical calculation of higher-order (beyond third-order) corrections are extremely complicated.  Therefore, it is significant to introduce a new high-accuracy analytical method for studying the phase transition between incompressible (MI or CDW) phases and compressible (SF and SS ) phases.

In this paper, we will use the GEPLT to study zero temperature phase diagrams of extended Bose-Hubbard model on optical square lattice  with two typical cases i.e., $ZV=0.1U$ and $ZV=1.5U$. The GEPLT is introduced by Wang et al. \cite{wang} and improved (some technical issues in Wang et al.'s work \cite{wang} were clarified) by  Lin et al. \cite{zhi-4}. The clarified GEPLT is a good method for investigating the quantum phase transition of bipartite superlattice systems (bipartite sublattice induced by superlattice structure) and the corresponding phase boundaries obtained by the GEPLT are in good accordance with the QMC simulations \cite{zhi-4}. Hence, we believe that the GEPLT can also be used to study the phase diagrams of extended Bose-Hubbard model on optical square lattice which is the other type of bipartite optical lattice systems caused by  nearest-neighbor repulsions $V$.

\section{the model and method}
We will consider the extended Bose-Hubbard model with an isotropic nearest-neighbor repulsion and the corresponding Hamiltonian reads
\begin{equation}
\hat{H}_{\mathrm{eBH}}=-t\!\sum_{\langle i,j\rangle}\!\hat{a}_{i}^{\dagger}\hat{a}_{j}+\!\sum_{i}\!\frac{U}{2}\hat{n}_{i}(\hat{n}_{i}-1)+V\!\sum_{\langle
i,j\rangle}\!\hat{n}_{i}\hat{n}_{j} -\mu\!\sum_{i}\hat{n}_{i}, \label{EBH}
\end{equation}
where $t$ is the hopping matrix elements of the bosons between
the nearest-neighbor sites $i$ and $j$, $\hat{a}_{i}^{\dagger}$ ($\hat{a}_{i}$) is the boson creation (annihilation) operator at site $i$, $\hat{n}_{i}=
\hat{a}_{i}^{\dagger}\hat{a}_{i}$\ is the particle number operator
on $i$-th site, $U$ denotes the on-site repulsion, $V$ is isotropic nearest-neighbor repulsions and $\mu$ is the chemical potential. In the atomic limit ($t=0$), it is easy to know that the ground states of systems have two different types. In other words, in the case $ZV>U$ the ground state is just $\rm{CDW}(n,0)$ for $(n-1)U<\mu\leq nU$,  in contrast case, the ground states vary between $\rm{CDW}(n,0)$ for $(n-1)(U+ZV)<\mu\leq (n-1)U+nZV$ and $\rm{MI}(n,n)$ for $(n-1)U+nZV<\mu\leq n(U+ZV)$ \cite{SC,Iskin,GGF1}.

\begin{figure}[h!]
\centering
\includegraphics[width=1.0\linewidth]{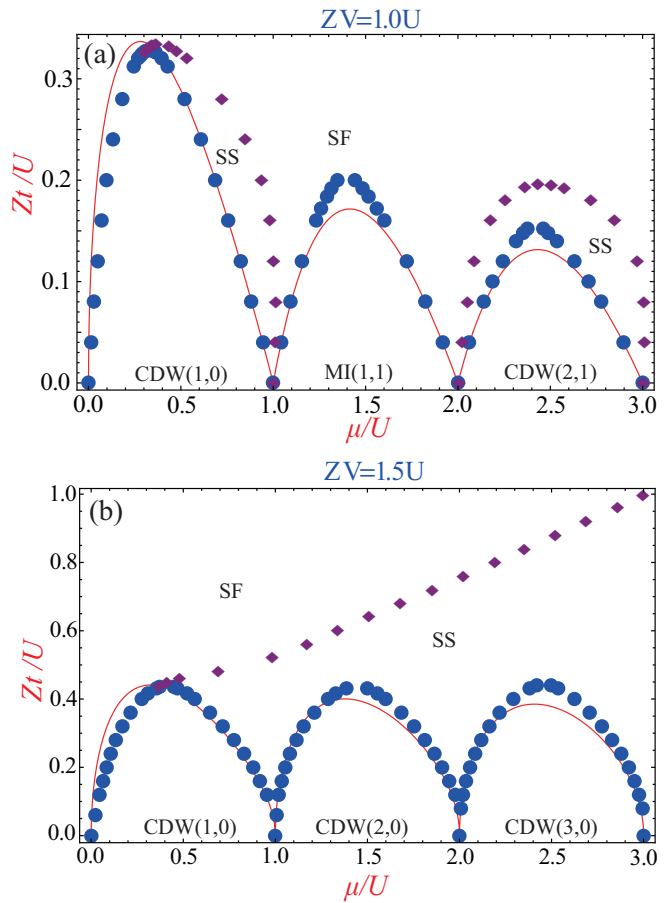}
\caption{(Color online)The phase diagram of extended Bose Hubbard model on optical square lattices with $ZV=U$(a) and $ZV=1.5U$ (b). The blue circles and purple rhombus, obtained by QMC  simulation, denote the boundaries of the insulating lobes and SF-SS respectively. The red line indicates the boundaries of insulating lobes, which are calculated by mean field method.}.
\label{QMC-MF}
\end{figure}

For the atomic limit, at the critical point $(ZV/U)_{c}=1$, the ground states are degenerate states. Specifically, $\rm{CDW}(n+1,n)$ [$\rm{MI}(n,n)$] is degenerate with the $\rm{CDW}(2n+1,0)$ [$\rm{CDW}(2n,0)$] \cite{Iskin}. Due to the quantum fluctuations, the critical point $(ZV/U)_{c}$ is slightly shifted to a lager value for finite hopping amplitude $t$ \cite{QMC11}. Therefore, the phase diagrams of the points $ZV=1.5U$ and $ZV=1U$ belong to two different types. At these points, the corresponding phase diagrams obtained by unbiased QMC simulations have been calculated \cite{QMC11} and it is shown in Fig. \ref{QMC-MF}, where blue circles, purple rhombus are QMC data and the red line is  the mean field results. From the Fig. \ref{QMC-MF}, it is obvious that the mean field results are smaller than the QMC simulations (exact results) except the $\rm{CDW}(1,0)$ lobe (having triple point on the lobe). Moreover, in the finite $t$ systems, the points $ZV=1.5U$ and $ZV=1U$ represent two different types, i.e., $ZV>(ZV)_{c}$ and $ZV<(ZV)_{c}$, and we can infer that the mean filed results always underestimate the QMC results at least in the cases of $ZV\neq (ZV)_{c}$. In other words, the results obtained by beyond mean field analytical method whether or not beyond the mean field results can be as an evidence to inspect that this analytical method is valid or invalid.

M. Iskin et al.'s work \cite{SC} shows that in a $d\rightarrow \infty $ dimensional hypercubic lattice, the extrapolated strong-coupling results for CDW-SS phase boundary do not always coincide with the mean field results which are exact for $d\rightarrow \infty$, but the extrapolated strong-coupling consequences for MI-SF phase boundary always coincide with the mean field solutions (see Fig.~3 in M. Iskin et al.'s work \cite{SC}). Moreover, for $ZV\gtrsim 0.7U$, M. Iskin et al. declared that the reason why the third-order strong-coupling perturbation theory is ineffective to obtain the CDW-SS phase boundary is that the fourth-order correction is essential. Unfortunately, if we use strong-coupling perturbation theory to calculate CDW-SS phase boundary for the case $ZV\gtrsim 0.7U$, as an example $ZV=1.5U$, there is an unphysical divergency [the denominator $U(n_{B}-n_{A}-1)+V(Zn_{A}-Zn_{B})$ in Eqs.~16 and 17 of M. Iskin et al.'s work \cite{SC} will be zero for $\rm{CDW}(n_{A}=2,n_{B}=0)$ ] which can not be eliminated by adding the fourth-order correction. In order to exclude the fourth-order correction effect, we can compare the third-order strong-coupling perturbation theory results with the mean filed results in a square lattice with $ZV=0.1U$. In Figs.~\ref{ZV01} (b) and (c), the third-order strong-coupling perturbation theory results (using Eqs.~16 and 17 of M. Iskin et al.'s work \cite{SC} ) are not always larger than mean field results. The above-mentioned argument implies that the strong-coupling perturbation theory is not always valid for calculating CDW-SS phase boundaries.

Therefore, we need a new analytical method which is effective  for obtaining the phase boundaries between incompressible (MI or CDW) phases and compressible (SF and SS ) phases in all parameters space of extended Bose Hubbard model. Fortunately, although the GEPLT is established for studying the phase transition of a superlattice or a multicomponent system \cite{wang,zhi-4}, under slight changing, the GEPLT can also be used to study the multipartite lattice system, where multipartite sublattice structure is caused by dipolar interaction.

\begin{figure}[h!]
\centering
\includegraphics[width=1.0\linewidth]{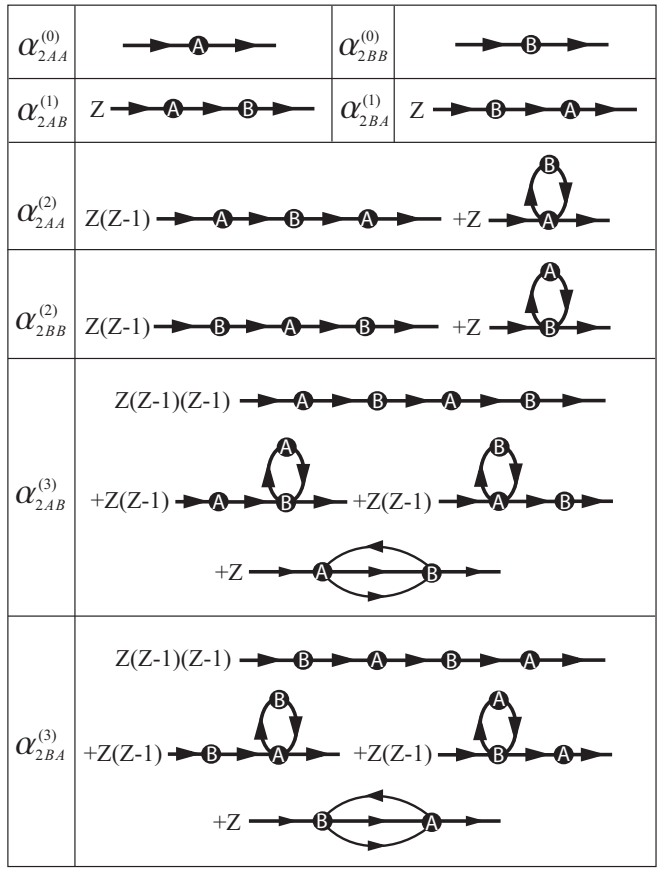}
\caption{The diagrams of the coefficients $\alpha^{(n)}_{2ij}$ for extended Bose Hubbard model on square lattice, where $Z=4$ is the coordination number of the systems.}
\label{table-1}
\end{figure}

The GEPLT has been developed \cite{wang,zhi-4} for studying the phase transition of  superlattice systems. For the multipartite lattice system caused by dipolar interaction, the GEPLT can also be set up and the corresponding process of establishing the GEPLT is similar to the process of establishing the GEPLT in superlattice systems \cite{wang,zhi-4}. We only need to replace Bose Hubbard model by extended Bose Hubbard model (specifically speaking, using $\hat{H}_{\mathrm{eBH}}$ instead of $\hat{H}$ in Eq. 5 of Lin et al.'s work \cite{zhi-4}) and then follow the same processes in Lin et al.'s work \cite{zhi-4}, thus the GEPLT, for studying the phase transition of extended Bose Hubbard model, is clearly established. The corresponding n-th order phase boundaries $t^{(n)}_{c}$ of extended Bose Hubbard model on an optical square lattice are given by
\begin{equation}
t^{(1)}_{c}=\frac{\sqrt{\alpha_{2AA}^{(0)}\alpha_{2BB}^{(0)}}}{\alpha_{2AB}^{(1)}},\label{mean-field}
\end{equation}
\begin{equation}
t^{(2)}_{c}=\frac{2\sqrt{\alpha_{2AA}^{(0)}\alpha_{2BB}^{(0)}}\alpha_{2AB}^{(1)}}{\alpha_{2AA}^{(0)}\alpha_{2BB}^{(2)}+\alpha_{2BB}^{(0)}\alpha_{2AA}^{(2)}},
\label{2-order}
\end{equation}
\begin{equation}
t^{(3)}_{c}=\frac{\alpha_{2AA}^{(0)}\alpha_{2BB}^{(2)}+\alpha_{2BB}^{(0)}\alpha_{2AA}^{(2)}}{2\sqrt{\alpha_{2AA}^{(0)}\alpha_{2BB}^{(0)}}\alpha_{2AB}^{(3)}},
\label{3-order}
\end{equation}
where $\alpha^{(n)}_{2ij}$ is perturbative coefficient for bipartite sublattice systems. Through Raylieigh-Schr\"{o}dinger perturbation expansion, these perturbative coefficients $\alpha^{(n)}_{2ij}$ can be calculated. Similarly to Feynman diagrams method that can greatly simplify the procedure of calculating, the perturbative coefficients $\alpha^{(n)}_{2ij}$ can also be represented by a set of diagrams to simplify the calculation. The corresponding diagrammatic representations have been formally introduced \cite{santos,zhi-1,wang}. Using the rule of diagrammatic representations \cite{santos,zhi-1,wang},  the first four orders of diagrammatic expressions of $\alpha^{(n)}_{2ij}$ are shown in the Fig.~\ref{table-1}. Moreover, the first-order phase boundary $t^{(1)}_{c}$ coincides with the mean-field result \cite{SC,Iskin}.
 \begin{figure}[h!]
\centering
\includegraphics[width=1.0\linewidth]{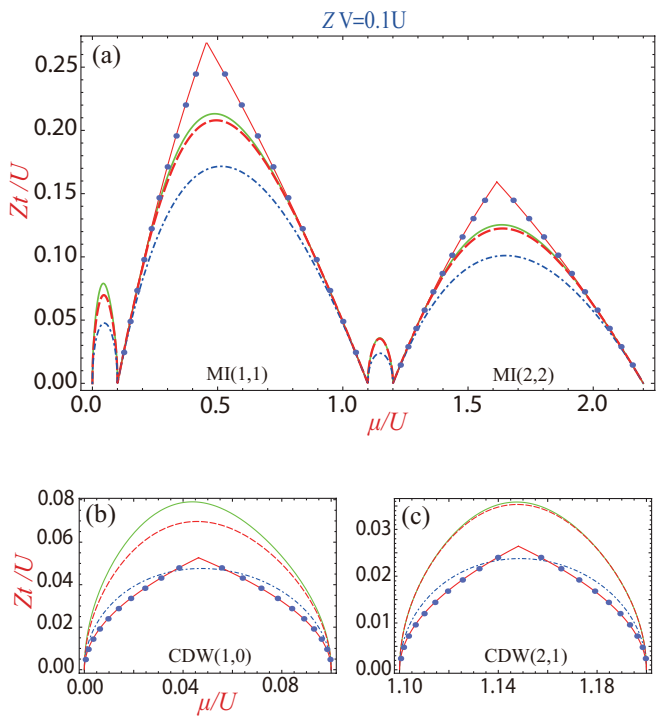}
\caption{(Color online)The phase diagram of ultracold Bose gases with isotropic nearest-neighbor repulsions $V$ in square lattice for the case of  $ZV=0.1U$.
The dot line is determined by the third-order strong-coupling perturbation theory. The other lines are  obtained by the GEPLT, such as the green solid line is the third-order calculation, the red dashed line is the second-order result, and the blue dot-dashed line is the first-order (mean-field) result. The results obtained by the GEPLT are in good agreement with the third-order strong-coupling perturbation theory calculations for MI lobes, but the GEPLT is better than the third-order strong-coupling perturbation theory for calculating CDW lobes. }
\label{ZV01}
\end{figure}

\section{The ground-state phase diagram}

\begin{figure}[h!]
\centering
\includegraphics[width=1.0\linewidth]{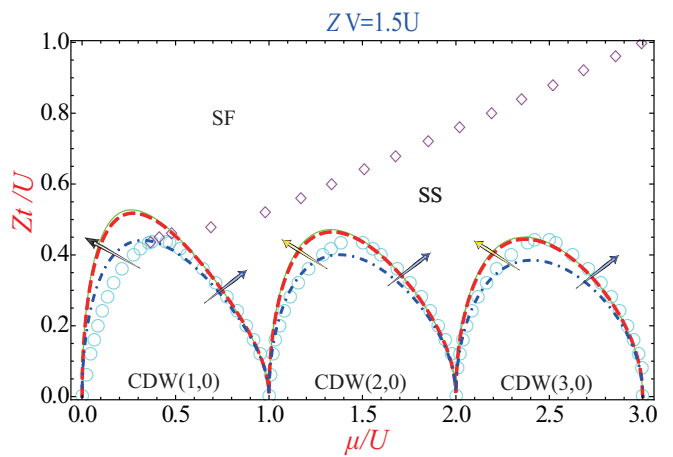}
\caption{(Color online)The phase diagram of ultracold Bose gases with isotropic nearest-neighbor repulsions $V$ in square lattice for the case of  $ZV=1.5U$.
Here the cyan circle are given by the QMC simulations and the lines are determined by the GEPLT, where the green solid line is the third-order calculation, the red dashed line is the second-order result, and the blue dot-dashed line is the first-order (mean-field) result.
Along the black arrow, the system will appear the first order phase transition, along the yellow arrow the system will appear the weak first-order phase transition and along the blue arrow the system will undergo the second-order phase transition.}
\label{phase-diagram-superlattice2}
\end{figure}
Using phase boundaries in Eqs.~(\ref{mean-field})-(\ref{3-order}), we can analytically obtain the first, second and third order phase boundaries between incompressible (MI or CDW) phases and compressible (SF and SS ) phases in all parameters space except $ZV=(ZV)_{c}$. Since the CDW-SS phase transition is
second-order phase transition for $ZV <U$ \cite{Gutzwiller2} and MI-SF phase transition is always second-order phase transition for all parameters space, thus the incompressible-compressible phase transitions are always second-order phase transition with $ZV <U$. Due to the fact that the GEPLT is only valid for second-order phase transition, the GEPLT can be used to calculate the phase boundaries of incompressible-compressible phase transitions of extended Bose Hubbard model on a square lattice with $ZV <U$.

According to  M. Iskin et al.'s argument, the fourth-order correction for the CDW lobes is negligible at point $ZV=0.1U$, thus we choose the point $ZV=0.1U$
as a representative point to calculate the phase boundaries between incompressible-compressible phases and to compare the phase boundaries between an incompressible and a compressible phase obtained by the third-order strong-coupling perturbation theory. From the Fig.~\ref{ZV01}, our analytical results are in good agreement with the third-order strong-coupling perturbation theory calculation for the edges of the MI lobes, moreover our analytical results are better than the third-order strong-coupling perturbation theory calculation for the tips of the MI lobes. For CDW lobes, since the boundaries obtained by the third-order strong-coupling perturbation theory are below the boundaries given by mean field method for lager region, it implies that the third-order strong-coupling perturbation theory is inaccurate.

At the case of $ZV=1.5U$, the ground state belongs to the other type, where the incompressible phase is only CDW phase.  Even if the fourth-order correction of strong-coupling perturbation theory is involved, strong-coupling perturbation theory is not always valid, since there is an unphysical divergency for $\rm{CDW}(2,0)$ phase. Thus, up to now, the GEPLT is the only possible analytical method which can be use to calculate the phase boundaries between CDW phase and SS phase. We have calculated the first three orders of the CDW-SS phase boundaries which are shown in Fig.~\ref{phase-diagram-superlattice2}. From the Fig.~\ref{phase-diagram-superlattice2}, it is clear that the parts (marking with blue arrow) of CDW lobes obtained by our analytical method are in quantitative agreement with QMC simulation, but there are significant differences for the other parts of CDW lobes (marking with black or yellow arrow) given by two different method. T. Ohgoe et al.'s work \cite{QMC11} shows that the type of phase transition for CDW-SS phase transition is dependent on the chemical potential $\mu$. More specifically, for the large $\mu$ part of each CDW lobe marked with blue arrow, the type of phase transition is second-order phase transition. But for the small $\mu$ part of each CDW lobe marked with black (yellow) arrow, the type of phase transition is first-order (weak first-order) phase transition. The reason why only the second-order phase transition CDW lobes given by our analytical method and QMC simulations are in excellent agreement with each other is that the GEPLT is only effective for second-order phase transition. Because CDW lobes for the second-order phase transition part given by our analytical method and QMC simulations coincide with each other, the GEPLT is beneficial to investigating the phase boundaries of second-order phase transition of extended Bose Hubbard model on a square lattice.

\section{Conclusion}
By comparing the mean-field results and QMC simulations for two different typical types, we infer that if the high precision analytical method is effective for extended Bose Hubbard model, the results obtained by this method must be greater than the mean field results. At one representative point $ZV=0.1U$, the phase boundaries between compressible phase and incompressible phase have been calculated by the GEPLT and the third-order strong-coupling perturbation theory. The corresponding results show that the third-order strong-coupling perturbation theory is invalid (valid) for obtaining the CDW (MI) lobes and the GEPLT is always valid for obtaining both CDW and MI lobes. Moreover, the GEPLT is better than the third-order strong-coupling perturbation theory for obtaining the tips of the MI lobes. At another representative point $ZV=1.5U$, even if the fourth-order correction is involved, the strong-coupling perturbation theory fails for obtaining the incompressible-compressible phase boundaries, because the phase boundaries Eqs.~(16) and (17) in reference \cite{SC} will be divergent for $\rm{CDW}(2,0)$ state. But the GEPLT is good for obtaining the part of CDW lobes with the second-order phase transition. The GEPLT is beneficial to studying the phase boundaries of second-order phase transition of extended Bose Hubbard model on a square lattice.

\section*{Acknowledgement}

Zhi Lin acknowledges Takahiro Ohgoe for providing the QMC data. Zhi Lin is supported by the Open Project of State Key Laboratory of Surface Physics in Fudan University (KF$2018\_13$).

\end{document}